\documentclass[preprint,showpacs,amsmath,amssymb,pra]{revtex4}

\usepackage{epsfig}
\usepackage{graphicx}
\newcommand {\um} {{\bf 1}}
\newcommand {\dr} {{\rm d}}
\newcommand{\bfrho}{\mbox {\boldmath $\rho$}}
\newcommand{\bbfrho}{\mbox {\boldmath $\widetilde\rho$}}
\newcommand{\bfPsi}{\mbox {\boldmath $\Psi$}}

\newcommand{\Ac}{\boldmath ${\cal A}$}
\newcommand{\BB}{\boldmath ${\cal B}$}
\newcommand{\Ec}{{\boldmath ${\cal E}$}}
\newcommand{\GG}{{\boldmath $\Gamma $}}
\newcommand{\HH}{{\boldmath ${\cal H}$}}
\newcommand{\II}{\boldmath ${\cal I}$}
\newcommand{\KK}{\boldmath ${\cal K}$}
\newcommand{\MM}{\boldmath ${\cal M}$}
\newcommand{\NN}{\boldmath ${\cal N}$}
\newcommand{\QQ}{\boldmath ${\cal Q}$}
\newcommand{\RR}{\boldmath ${\cal R}$}
\newcommand{\Sc}{\boldmath ${\cal S}$}
\newcommand{\TT}{\boldmath ${\cal T}$}
\newcommand{\VV}{\boldmath ${\cal V}$}
\newcommand{\WW}{\boldmath ${\cal W}$}
\newcommand{\bff}{\boldmath ${\cal F}$}
\newcommand{\bbff}{\boldmath ${\widetilde {\cal F}}$}
\begin{document}
\title{Multi-Channel Inverse Scattering Problem on the Line: 
Thresholds and Bound States}
\author{
     M. Braun, S. A. Sofianos, H. Leeb\footnote{On leave 
from Atominstitut der \"Osterreichischen 
     Universit\"aten, Technische Universit\"at Wien,
     Wiedner Hauptstra\ss e 8-10, A-1040 Wien, Austria}}
\affiliation{Physics Department, University of South Africa,
     Pretoria 0003, South Africa}
\date{\today}

\pacs{ 03.65.Nk}

\begin{abstract}
We consider the multi-channel inverse scattering problem 
in one-dimension in the presence of thresholds and bound  
states for a potential of finite support. Utilizing the Levin 
representation, we derive the general Marchenko integral equation 
for N-coupled channels and show that, unlike to the case of the 
radial inverse scattering problem, the information on the bound state
energies and asymptotic normalization constants can be inferred 
from the reflection coefficient matrix alone. Thus, given this matrix, 
the Marchenko inverse scattering procedure can provide us with a 
unique multi-channel potential. The relationship to supersymmetric 
partner potentials as well as possible
applications are discussed. The integral equation has been implemented
numerically and applied to several schematic examples showing the
characteristic features of multi-channel systems.
A possible application of the  formalism to technological problems
is briefly  discussed.
\end{abstract}

\maketitle

\section{Introduction}

The first attempt to extend the single-channel inverse scattering problem 
(ISP) on the line \cite{Kay60,Marchenko63,Chadan89,Lamb81,Ghosh91} to
a wider class of $N\times N$ coupled-channel potentials was made by
Wadati and Kamijo \cite{Wadati74} about $25$ years ago. They derived a 
Marchenko equation associated with the $N\times N$ Schr\"odinger 
equation on the entire line $(-\infty<x<\infty)$. The problem has also 
been discussed in detail by Calogero and Degasperis
\cite{Calogero77}. In these investigations the presence of threshold 
energies was not included and, most importantly, practical aspects of 
the implementation of the solution of the ISP were not considered. 
Twenty years later, the coupled-channel problem was taken further 
with the inclusion of thresholds \cite{Sofianos97} and 
numerical solution of the corresponding Marchenko equation.

The lack of progress can be attributed to several reasons. 
Firstly, the solution of the ISP presupposes the knowledge of the full
reflection coefficient (moduli and phases) at all incident momenta $q$.
This requirement can not be  easily fulfilled, especially due to the 
missing phase information in standard experiments, 
a diffiuclty which is similar to   
the well known and longstanding phase problem in diffraction analysis
\cite{Cowley75,Burge76,Klibanov92}. Secondly, a profile with an 
unlimited support 
leads to tough numerical questions because the highly oscillatory 
behavior of the reflection coefficient makes the numerical procedure 
cumbersome and unstable. A third difficulty in the presence of bound
states is the determination of the bound state
normalization constants for the  various channels which are crucial 
for a unique reconstruction of the underlying coupled-channel 
potential. 

Coupled-channel inverse scattering techniques in one dimension have 
been considered and numerically implemented for specular reflection of 
polarized neutrons from plane surfaces of magnetized samples 
\cite{Kasper99}. These problems contain neither thresholds nor bound 
states and can be treated by the available Marchenko equations. The 
formulation of coupled-channel inverse scattering techniques taking 
simultaneously into account thresholds and bound states would be a 
valuable progress, specifically for the design of semiconductor 
quantum devices and for the synthesis of quantum heterostructures with 
specific spectral properties \cite{QD99}. In this context the spectral 
design of systems with specific bound states embedded in the continuum 
would be of great interest for application.

In the present work we focus on the formulation of equations and
numerical procedures for the solution of the coupled-channel inverse 
scattering problem on the line including thresholds. We also take into 
account bound states lying energetically below the thresholds of all
channels, but exclude those embedded in the continuum of any channel. 
Furthermore, we point out that the neglect of bound states in the
input of the inversion procedure results in a supersymmetric partner
potential with the same reflection properties, but completely
different spatial shape. To test the formulated inverse scattering 
procedure and their numerical implementation,  model examples 
are considered. In these models the reflection coefficient has 
been evaluated using known profiles and employed as input in the 
inverse scattering equations 
to recover the original scattering profile. The agreement between the 
reconstructed and the original profile is an important measure for 
the quality of the inverse scattering procedure and a criterion for 
the reliability of the numerical methods.

In Sect. II we present the formalism and give the essentials for the
derivation of the generalized Marchenko equation. Technical details 
concerning the evaluation of the corresponding integrals and the 
treatment of bound states are outlined in the Appendix \ref{app:A}. 
Furthermore, we sketch briefly the formalism of supersymmetric quantum 
mechanics for coupled-channel systems with thresholds and the 
construction of supersymmetric partner potentials.
The feasibility of the developed formalism is demonstrated in 
Sect. III, where we reconstruct successfully coupled-channel
potentials from simulated reflection and bound state data for various 
two-channel systems. Examples with and without thresholds and bound
states are considered. In Sect. IV we summarize our conclusions
and discuss potential applications of this inverse scattering 
technique.    

%
%
\section{Formalism}
%
\subsection{The direct problem}
The  multi-channel Schr\"odinger equation in one dimension has
the form (in appropriate units)
\begin{equation}
       \left( -\frac{d^2}{dx^2} + \mbox{\boldmath ${\cal V} $}(x)+
    \mbox{\boldmath ${\cal E} $}\right) 
    \mbox{\boldmath ${\Psi }$} =  
    k^2 \mbox{\boldmath ${\Psi }$} \,, 
\label{ccs}
\end{equation}
where {\VV }$(x)$ is assumed to be a real symmetric 
$N\times N$ potential matrix of finite support with matrix elements 
$V_{ij}(x)$, $i,j = 1, \cdots , N$. Without loss of generality we 
assume for convenience that {\VV }$(x)$ vanishes on the entire
negative $x$-axis. The solution \bfPsi {} is a $N\times N$ matrix 
whose columns are formed by the $N$ linearly independent solution 
vectors of (\ref{ccs}) at an incident momentum  $k$ and \Ec {} is a 
diagonal matrix containing the threshold energies $\epsilon_i$, 
$i = 1, \cdots , N$. We assume that the threshold energies,
$\epsilon_i$ are arranged in increasing order with the lowest (in
most cases the entrance channel) set to zero, $\epsilon_1=0$. The
special case of no thresholds, {\Ec }$=0$, has already been considered 
by Wadati and Kamijo \cite{Wadati74} in 1974. Here, we focus on the
general case and allow the occurrence of non-vanishing thresholds,
$\epsilon_j>0$ for $j=2,3,\dots ,N$.

The free solutions  are given by 
$\exp (\pm i${\mbox{\KK }}$x )$,  where  now {\mbox{\KK }} is given by
$$
     \mbox{\KK }^2 = k^2\um- \mbox{\Ec} \, ,
$$
$\um $ being the $N$-dimensional unit matrix. The matrix \mbox{\KK } is
diagonal and is defined on the physical sheet of the Riemann surface 
for the momentum variable $k$. This sheet has an $(N-1)$-fold branch 
cut on the real axis on the upper rim of which $k_1 =k $, and the 
other diagonal elements  of \mbox{\KK } are defined as follows
\begin{equation}
    k_j=
    \begin{cases}
      +\sqrt{k^2-\epsilon_j}  & {\rm for} \ \  k \ge \sqrt{\epsilon_j} \cr
      +{\rm i}\sqrt{\epsilon_j-k^2}  & {\rm for }\   
                \       \vert k \vert <   \sqrt{\epsilon_j}\cr
       -\sqrt{k^2-\epsilon_j} & {\rm for} \ \   k \le -\sqrt{\epsilon_j}\cr
\end{cases}
\label{kjj}
\end{equation}
for  $j = 2, \cdots, N$.

Similarly to the one-channel case we introduce sets of linearly 
independent matrix Jost solutions {\bff}$_\pm (k,x)$, 
{\bbff}$_\pm (k,x)$ defined by the boundary conditions
\begin{eqnarray}
      \lim_{x\to\pm\infty}\exp(\mp{\rm i}\mbox{\KK } x)\mbox{\bff}_\pm(k,x)
   &=&\um\\
   \lim_{x\to\pm\infty}\exp(\pm{\rm i}\mbox{\KK } x)\mbox{\bbff}_\pm(k,x)&=&\um
\, .
\label{asyma}
\end{eqnarray}
The boundary condition {\bbff}$_-(k,x)$, for example, implies
 that each solution vector has at $x\to -\infty $ 
only in one channel an incoming wave from the left ($e^{ik_jx}$) and 
none in all other channels. It is therefore obvious that
{\bbff}$_-(k,x)$ will provide the incoming component of the physical 
solution {\bfPsi}$_L (k,x)$ for incidence from the left. Because the 
Jost solutions form a complete set we can write the physical solution 
of (\ref{ccs}) for left and right incidence in terms of 
{\bff}$_\pm (k,x)$ and {\bbff}$_\pm (k,x)$,
\begin{eqnarray}
    \mbox{\bfPsi}_L (k,x) & = & \mbox{\bff}_+ (k,x) \mbox{\TT}_L(k)
\nonumber
\\
      &=& \mbox{\bbff}_- (k,x) + \mbox{\bff}_- (k,x) \mbox{\RR}_L(k)
\, , \label{physL}\\
\nonumber
\mbox{\bfPsi}_R (k,x) & = & \mbox{\bff}_- (k,x) \mbox{\TT}_R(k)\\
      & =& \mbox{\bbff}_+ (k,x) + \mbox{\bff}_+ (k,x) \mbox{\RR}_R(k) 
\, . \label{physR}
\end{eqnarray}
Here, {\RR}$_{L,R}(k)$ and \mbox{\TT}$_{L,R}(k)$ are the reflection
and transmission coefficient matrices for left (L) and right (R) 
incidence of the beam, respectively.

In the absence of thresholds the members of each pair of solutions
are related by replacing $k$ by $-k$, i.e. 
{\bbff}$_\pm (k,x)=${\bff}$_\pm (-k,x)$ (cf. \cite{Wadati74}). 
In the presence of thresholds, however, this relation does not 
hold anymore for $k^2 < \epsilon_N$ because of the $N$-fold 
connectivity of the $k$-plane \cite{Sofianos97}. Considering the 
symmetry of the transformation (\ref{kjj}) and of the Schr\"odinger 
equation (\ref{ccs}) it is obvious that the identities
\begin{equation}
   \mbox{\bff}_\pm (-k,x) = \mbox{\bff}_\pm^* (k,x) \quad \mbox{and}
    \quad \mbox{\bbff}_\pm (-k,x) = \mbox{\bbff}_\pm^* (k,x)
\label{fmk}
\end{equation}
are also satisfied in the presence of thresholds. A direct
consequence of Eq. (\ref{fmk}) is the symmetry property of the
reflection and transmission matrices for real $k$,
\begin{equation}
	\mbox{\RR}_{R,L}(-k) = \mbox{\RR}^*_{R,L}(k) \quad \mbox{and}
	\quad \mbox{\TT}_{R,L}(-k) = \mbox{\TT}^*_{R,L}(k) \, .
\end{equation}
Introducing the matrix generalized Wronskian relation,
\begin{eqnarray}
\nonumber
    W[\Psi (k,x),\Phi(k,x)]&\equiv&
    \Psi^T(k,x) \,\left( \frac{{\rm d}}{{\rm d}x}\Phi(k,x)\right)\\
    &-&
    \left( \frac{{\rm d}}{{\rm d}x}\Psi^T(k,x)\right)\ \Phi (k,x)
\, ,
\end{eqnarray}
where $\Psi $ and $\Phi $ are solutions of the coupled-channel
Schr\"odinger equation (\ref{ccs}) and $T$ denotes transposition,
one obtains 
\begin{equation}
      \frac{d}{dx} W[\Psi (k,x),\Phi (k,x)]=0 \, .
\end{equation}
This means that the Wronskian of two solutions of Eq. (\ref{ccs})
is constant on the entire $x$-axis. Specifically, it vanishes for two
linearly dependent solutions. Evaluating the Wronskian 
$W[\Psi_L,\Psi_L]=0$ in the limit $x\to -\infty $ and using the relation
of $\Psi_L(k,x)$ given in Eq. (\ref{physL}), yields 
\begin{equation}
        \mbox{\RR}_L^T \mbox{\KK} = \mbox{\KK} \mbox{\RR}_L \, .
\label{rever}
\end{equation}
This relation clearly indicates the effect of the thresholds on the
symmetry of the reflection matrix.
\subsection{Inverse Scattering Equation}
The determination of the potential matrix, {\VV}$(x)$, from the
knowledge of the reflection matrix, {\RR}$_{L}(k)$ ({\RR}$_R(k)$), is 
known as inverse scattering problem and its solution can be obtained
via integral equations, often referred as Marchenko equations, for 
several quantum systems. Recently, an integral equation for 
one-dimensional coupled-channel systems with thresholds has been
presented \cite{Sofianos97}. The derivation did not include 
the presence of bound states which are essential features of
realistic quantum systems. Here, we focus on the ISP of 
coupled-channel systems in one dimension including thresholds 
and bound states. In the following the essential aspects of the 
derivation are outlined.

The integral equation for the solution of the ISP associated with
Eq. (\ref{ccs}) is most easily obtained via the Levin representation
\cite{Chadan89,Ghosh91} of {\bff}$_\pm (k,x)$,
\begin{eqnarray}
    \mbox{\bff }_+(k,x)&=&{\rm e}^{+i\mbox{\KK}x}+ \int^{+\infty}_x\,
             dz \ \mbox{\BB}_+(x,z){\rm e}^{+i\mbox{\KK}z} 
\label{Levin1} \\
    \mbox{\bff }_-(k,x)&=&{\rm e}^{-i\mbox{\KK}x}+ \int_{-\infty}^x\,
             dz \ \mbox{\BB}_-(x,z){\rm e}^{-i\mbox{\KK}z}. 
\label{Levin2}
\end{eqnarray}
Inserting these expressions into Eq. (\ref{ccs}) leads after some
algebraic manipulations to differential equations for the
transformation kernels {\BB }$_\pm (x,y)$,
\begin{equation}
\left( 
     \frac{\partial^2}{\partial x^2}-\frac{\partial^2}{\partial y^2}
     \right) \mbox{\BB}_\pm (x,y) = \mbox{\VV }(x) \mbox{\BB}_\pm (x,y)
+ \left[ \mbox{\Ec},\mbox{\BB}_\pm (x,y)\right] \, ,
\label{gder}
\end{equation}
where we have introduced the commutator [{\Ac},{\BB}]={\Ac}{\BB}-{\BB}{\Ac}.
The transformation kernels are related to the potential via
\begin{equation}
\mbox{\VV}(x) = -2\frac{d}{dx}\mbox{\BB}_+(x,x^+)\, ,\quad
\mbox{\VV}(x) = 2\frac{d}{dx}\mbox{\BB}_-(x,x^-)
\end{equation}
and satisfy the boundary conditions
\begin{equation}
\lim_{x,y\to \pm \infty } \mbox{\BB}_\pm (x,y) = 0
\, .
\label{Bbc}
\end{equation}
Thus for $x=y$,
\begin{equation}
\mbox{\BB}_-(x,x)=\frac{1}{2}\int_{-\infty}^x dz \mbox{\VV}(z) 
\, .
\label{rgbound}
\end{equation}
The above partial differential equations together with the boundary 
conditions (\ref{Bbc}) and (\ref{rgbound}) constitute a Goursat
problem \cite{Ghosh91} of a generalized nature.

Because of the restriction placed on the potentials, the existence of 
solutions for the partial differential equation (\ref{gder}) 
under the above boundary conditions, can be easily shown as follows:
A change of variables to 
        $u=\frac{1}{2}(x-y)$ and $v=\frac{1}{2}(x+y)$ 
transforms  (\ref{gder}) into 
\begin{equation}
        \frac{\partial^2  {\bf B}_-}{\partial u \partial v} 
        = \mbox{\VV}(u+v) \mbox{\BB}_- +
        \left[ \mbox{\Ec}, \mbox{\BB}_- \right]\,;
\label{vderuv}
\end{equation}
A formal iterative  solution of this partial differential equation 
involves repeated integrations over the potential. Since we imposed
the condition of finite support on {\VV} this series will converge, 
due to the triangular nature of domains over which the integrations 
take place. 

The corresponding Marchenko integral equation is obtained by multiplying
Eq. (\ref{physL}) from the right with
\begin{equation}
\frac{1}{2\pi}\exp\left(-\mbox{i\KK} y \right) 
\frac{\mbox{d\KK}}{{\rm d}k}
=\frac{1}{2\pi}\exp\left(-\mbox{i\KK} y \right) \mbox{\KK}^{-1} k
\, ,
\label{deriv}
\end{equation}
substituting the Levin representations for {\bff}$_-(k,x)$ and 
{\bbff}$_-(k,x)$ on the right hand side and integrating over $k$.
Restricting ourselves to systems which have only bound states at 
negative energies ($k^2=q_\alpha^2<0$, $\alpha =1,\dots ,N_b$) and 
assuming that no bound states are embedded in the continuum, we can 
evaluate the relevant integrals (see Appendix \ref{app:A}). The 
result can be cast into the form
\begin{equation}
\mbox{\BB}_-(x,y)\,+\,\mbox{\bfrho}_-(x,y)\,+\,\int_{-\infty}^{x}\,
          {\rm d}z \mbox{\BB}_-(x,z)\mbox{\bfrho}_-(z,y)\,=0\,. 
\label{Mar1m}
\end{equation}
where $x>y $. The input kernel of the above integral equation is given by
\begin{equation}
  \bfrho_-(x,y)= \bbfrho_-(x,y) + \bbfrho^{(b)}_-(x,y)
\label{rho1mfb}
\end{equation}
with
\begin{eqnarray}
\bbfrho_-(x,y)&=&\frac{1}{\sqrt{2\pi}} \int_{-\infty}^{+\infty}\,
      {\rm d}k\,{\rm e}^{-i\mbox{\KK} x}  \mbox{\RR}_L(k)\,
    {\rm e}^{-i\mbox{\KK} y}\, ,
\label{rhoscat}\\
\bbfrho^{(b)}_-(x,y)&=&
   -{\rm i} \sum_{\alpha=1}^M {\rm e}^{-i\mbox{\QQ}_\alpha x}
    \mbox{\MM}_\alpha {\rm e}^{-i\mbox{\QQ}_\alpha y}
    \mbox{\QQ}^{-1}_\alpha q_\alpha \, ,
\label{rhobound}
\end{eqnarray}
and {\QQ}$_\alpha = $\mbox{\KK }$(q_\alpha )$.

\subsection{Determination of the Input Kernel}
An important step for the application of the inverse scattering 
procedure is the determination of the input kernel {\bfrho}$_-(x,y)$
from the scattering data. Using the relation 
{\RR}$_L(k)=${\RR}$_L^*(-k)$ one can immediately evaluate the 
contribution of the continuum spectrum, {\bbfrho}$_-(x,y)$.
The bound state contribution, {\bbfrho}$^{(b)}_-(x,y)$ requires the 
knowledge of the bound state energies $q^2_\alpha$ and the corresponding
 asymptotic normalization matrices {\MM}$_\alpha$. If these quantities 
are obtained from simulated data, assuming a given potential {\VV}$(x)$, 
then the inversion procedure should yield the original potential. In 
spectral design, however, these quantities are not available except 
when one is interested to have a bound state at a specific energy.
In such a case arbitrary values of {\MM}$_\alpha$, may result in
a rather extended profile which is of limited interest because it 
can not easily be materialized. 

To ensure that the inverse scattering procedure provides us with a 
potential {\VV}$(x)$ which vanishes for $x<0$ we must look for the 
required characteristics of {\bfrho}$_-(x,y)$. From the Levin 
representation of {\bff}$_-$, Eq. (\ref{Levin2}) it follows that 
{\BB}$_-(x,y)$ vanishes identically for $x<0$. Because of
Eq. (\ref{Mar1m}), the input kernel $\bfrho_-(x,y)$ must also vanish 
for $y<x<0$ and Eq. (\ref{rho1mfb}) reduces for $x<0$ to the identity
%
\begin{eqnarray}
      \bbfrho_-(x,y)&=& -\bbfrho^{(b)}_-(x,y)\\
& =& 
      {\rm i}\sum_{\alpha=1}^{\infty}  
      \exp(-i\mbox{\QQ}_\alpha x) \mbox{\MM}_\alpha 
      \exp(-i\mbox{\QQ}_\alpha y) \mbox{\QQ}_\alpha^{-1}q_\alpha 
\nonumber
\label{rhotil} 
\end{eqnarray}
where $y<x<0$. 
This relation is formally equivalent to the one-channel case
\cite{Braun95} and expresses the fact that the potential {\VV}$(x)$ can
be forced to vanish for $x<0$ if there is a series of bound states with 
binding energies $q_\alpha^2$ and normalizations {\MM}$_\alpha $ which 
compensate {\bbfrho}$_-(x,y)$. 

Because of the peculiar spatial form of the contribution, it is obvious 
that those bound states which are energetically closest to threshold 
determine the behavior of the potential at large negative
$x$-values. Thus for practical applications, we firstly assume to
have a finite number, $N_b$, of bound states, and restrict ourselves 
to those energetically closest to threshold. With a fixed number $N_b$ 
of bound states, it is then possible to deduce the bound state
parameters solely from the knowledge of {\RR}$_L$ via the 
nonlinear Eq. (\ref{rhotil}). We may solve this equation in two steps. 
Since the diagonal elements of {\BB}$_-(x,y)$ only depend on $x+y$, 
a one dimensional fit to the sum of exponentials will produce the 
$q_\alpha$-values and the diagonal elements  of  {\MM}$_\alpha$. 
The knowledge of the $q_\alpha$ values is then sufficient to obtain 
the remaining non diagonal elements of the {\MM}$_\alpha$. Therefore, 
the kernel ${\bfrho}_-(x,y)$ can be recovered and the determination 
of the potential can be achieved via the Marchenko equation. 
It should be noted that in the presence of thresholds, the input
kernel $\bfrho_-(x,y)$ depends on the variables $x$ and $y$
separately, while in the case of no thresholds it depends on the 
sum $x+y$. Furthermore, the input kernel has the symmetry property 
{\bfrho}$_-(x,y)=${\bfrho}$^T_-(y,x)$ because of the symmetry, 
Eq. (\ref{rever}), which implies
{\RR}$_L$\mbox{\KK }$^{-1}=$\mbox{\KK }$^{-1}${\RR}$_L^T$.

\subsection{Supersymmetric Partners}
The omission of the contribution of the bound state spectrum in 
{\bfrho}$_-(x,y)$ leads via the inverse scattering procedure to a 
potential which generates the same reflection coefficient but does 
not sustain any bound state. This feature is characterizing 
phase-equivalent partner potentials which can be obtained easily
via techniques of supersymmetric (SUSY) quantum mechanics 
\cite{Witten81,Baye87,Sukumar85}. The corresponding so-called SUSY 
transformations are based on the factorization method \cite{Infeld51}
which has been formulated to coupled-channel systems including 
thresholds \cite{Amado87,Sparenberg97,Leeb00}. A compact outline of 
the extension to coupled-channel systems is sketched in the following.

In the factorization method the Hamiltonian of the coupled
Schr\"odinger equation (\ref{ccs}) is written in the form
\begin{equation}
-\frac{d^2}{dx^2}+\mbox{\VV}_0(x) = {\hat A}_0^+ {\hat A}_0^- 
- \mbox{\Ec} + {\bar q}^2 \, ,
\label{H0}
\end{equation}
where the factorization energy, ${\bar q}^2$, is smaller or equal to 
the energy $q_1^2$ of the lowest bound state. The index $0$ of 
{\VV}$(x)$ indicates that it is the original potential before any
transformation. The factorization operators ${\hat A}_0^\pm $ are
given in terms of the superpotential {\WW}$_0(x)$,
\begin{equation}
{\hat A}_0^\pm = \pm \frac{d}{dx} + \mbox{\WW}_0(x)\, ,
\label{Apm}
\end{equation}
which satisfies the nonlinear differential equation
\begin{equation}
\frac{d}{dx} \mbox{\WW}_0 + \mbox{\WW}^2_0 = 
\mbox{\VV}_0+\mbox{\Ec}-{\bar q}^2 \, .
\end{equation}
It is straightforward to show that for any solution $\Psi_0(k,x)$ of
the original Schr\"odinger equation (\ref{ccs}), the transformation
${\hat A}_0^- \Psi_0(k,x)$ leads to a solution at the same energy
$k^2$ of the coupled Schr\"odinger equation with the potential
\begin{equation}
\mbox{\VV}_1 = \mbox{\VV}_0 - 2\frac{d}{dx}\mbox{\WW}_0
\, .
\label{V1}
\end{equation}
If we choose the factorization energy ${\bar q}^2=q_1^2$, one is able 
to eliminate the ground state from the spectrum of the original
Hamiltonian (cf. \cite{Amado87,Sparenberg97,Leeb00}). The reflection
matrix {\RR}$_1(k)$ associated with the transformed potential
has changed, however,
\begin{equation}
     \mbox{\RR}_1(k) = (\mbox{\WW}_0^-+i\mbox{\KK}) \mbox{\RR}_0(k)
      (\mbox{\WW}_0^- - i \mbox{\KK})^{-1}
\, .
\end{equation}
Here, {\WW}$_0^\pm $ are the boundary values of {\WW}$_0$ at
$x\to \pm $, respectively. 

To restore the same reflection matrix {\RR}$(k)$ a second SUSY
transformation at the same energy ${\bar q}^2$ must be performed
\cite{Amado87,Sparenberg97} using the boundary values
{\WW}$_0^-=-${\WW}$_1^-= -i${\QQ}$_1$ \mbox{\Sc}. Here {\Sc} is
a diagonal matrix containing $-1$ in the first $M$ rows 
 and $+1$ in the remaining $N-M$, where $M\leq N$ is the 
degeneracy of the ground state of the original system. 
%
\section{Examples}
As a demonstration of the coupled-channel inverse scattering
equations, derived above, we consider several schematic examples 
which exhibit the specific features of coupled-channel systems. 
For the numerical implementation of the integral equation 
(\ref{Mar1m}) we follow similar  techniques as  outlined in 
the appendix  of Ref. \cite{Sofianos97}.
 
First we consider the case of a two-channel system without  a
threshold and a  bound state. We choose the potential {\VV}$(x)$
to have different $x$-dependence in the various matrix elements.
We take a Gaussian shape for $V_{11}(x)$ with the parameters 
     $ V_0=0.1$,
     $ b=4$,
     $ c=1.8$
and a two layer repulsive profile ($N=2$) for $V_{22}(x)$ with 
the parameters 
     $a=0.01$,
     $x_0=0.5$
     $V_1=0.08$,
     $x_1=2.7$,
     $V_2=0.05$  
     $ x_2=4.0$.
 The off-diagonal elements
$V_{12}(x)=V_{21}(x)$ are given by an $n_s=3$ sea-saw potential with
the parameters $V_0=0.075$,
               $x_\ell=1.2$,
               $x_s=0.75$,
               $ s_1=+1$, and
               $s_2=s_3=-1$.
A definition of the shapes and the parameters is given in Appendix 
\ref{app:B}. With this potential we solved Eq. (\ref{ccs}) and 
evaluated {\RR}$_L(k)$ up to $k_{max}=12$. Using this 
{\RR}$_L(k)$-values we reconstructed the potential {\VV}$(x)$ via 
the procedure given in Eqs. (\ref{Mar1m}) to (\ref{rhotil}). The 
reconstructed potential matrix elements are displayed in Fig. 1 
together with the original ones. It is seen that the 
reproduction of the original potentials is excellent.
%
%

Next we consider the case where thresholds are present. Again we
use an input potential {\VV}$(x)$  which is chosen
differently to the previous case in order to demonstrate the ability
of the algorithm to deal with rather different situations. The
potential chosen is a Gaussian with parameters 
      $ V_0=0.15$,
      $ b=9$,
      $c=1.8$
for $V_{11}(x)$; a one layer repulsive profile ($N=1$) with
      $a=0.05$,
      $x_0=1.0$,
      $V_1=0.20$,
      $x_1=2.8$
for $V_{22}(x)$;  and Gaussian potentials with
      $ V_0=0.12$,
      $ b=9$,
      $c=2.2$
for $V_{12}(x)=V_{21}(x)$. A threshold energy 
of $\epsilon=0.025$ is assumed in the second channel. The reconstruction
from evaluated {\RR}$_L(k)$ data up to $k_{max}=12$ are shown in
Fig. 2
and are once again in almost perfect agreement
with the original. It is interesting to note here that although
all elements of {\VV}$(x)$  are repulsive, the coupling of channels
may lead to bound states.
In the present example an increase of the strength of the coupling
potential e.g. from $V_0=0.12$ to $V_0=0.5$ results in a bound state
at $E_b=-0.00053$.  
This is in agreement with the observations made in Ref. \cite{Newton}.
%
%

An example of a two-channel system with a bound state, but without
threshold is shown in Fig. 3.
The potential is chosen
to be of Gaussian form with 
       $V_0=0.15$, 
       $ b=1.5$,
       $c=2.2$ 
for $V_{11}$; a one-layer profile with
       $a=0.1$,
       $x_0=1.0$, 
       $V_1=-0.1$
and    $x_2=3.3$,  
for $V_{22}$. The off-diagonal profiles
$V_{12}(x)=V_{21}(x)$ are given by a sea-saw potential with 
      $V_0=0.1$, 
      $x_\ell=1.5$, 
      $x_s=0.70$, and
      $n_s=2$,
with $s_1=s_2=+1$. The system sustains a bound state
at $E_b=-0.01558$  with
       $M_{11}=-0.0065$,
       $M_{12}=0.0216$, 
       $M_{21}=0.0216$, and 
       $M_{22}=-0.0711$.
As expected, the off-diagonal asymptotic normalization constants
are, in this case, the same. As one can see from Fig. 3, 
the  reproduction of the potential is, for all practical 
purposes, perfect.
%
%

It is interesting to consider next the simultaneous existence
of thresholds and bound states. For this we consider the previous
example with a threshold $\epsilon= 0.01$; the rest of the input 
data remaining the same.
The presence of the threshold generates a bound state
at   $E_b= -0.0068$ with
       $M_{11}=-0.0104$,
       $M_{12}=0.0400$, 
       $M_{21}=0.0257$, and 
       $M_{22}=-0.1031$.
The  reproduction of the potential is, once more, excellent.

As a final example, we consider the SUSY transformations.
Setting the asymptotic normalization constants equal to zero results, as
expected, in  supersymmetric partner profiles displayed in Fig. 4 
which do not sustain a bound state and
look considerably different from the original ones.


It should be noted here that  an insertion of a bound state
at an arbitrary energy results in different values for the
asymptotic normalization constants. When these constants
are evaluated via (\ref{rhobound}), then the inversion procedure
provides us always with the original interaction.

\section{Conclusions}
%
We have studied the inverse scattering problem on the line
including thresholds and derived the corresponding integral equation
of Marchenko type. Specific care was taken for the integrations 
on the physical sheet of the Riemann surface of the momentum variable 
$k$ because of the $(N-1)$-fold branch cut as discussed in Sect. IIA. 
We found that these branch cuts do not generate additional problems 
as long as we deal with systems sustaining only true bound states.
The integration contour is well defined and the integrand 
satisfies all conditions required for the application of
Cauchy's integral formula. In this way the resulting Marchenko
equation can be applied when threshold and bound states
are  simultaneously present in the system.

Much emphasis was given to the application of the method
which implies the numerical implementation as well as the
generation of the input kernel {\bfrho}$_-(x,y)$ from 
scattering data. Following the techniques outlined in Ref.
\cite{Sofianos97} we  achieved almost perfect numerical
reconstructions as it can be seen from the examples given
in Sect. III.

Furthermore, we considered the bound state contribution
to {\bfrho}$_-(x,y)$ which depends on the bound state energies
and the so-called asymptotic normalization constants, where the
latter are usually not accessible to experiment. We pointed out
that the neglect of the bound state contribution results in a
supersymmetric partner potential and leads to completely different
profiles which, nevertheless, generate the same reflection
coefficient matrix. Thus, the absence of bound state data 
may lead to wrong conclusions concerning the profiles.

 We proposed a method
for the determination of  bound state data with regard to the 
design of quantum devices with pre-designed reflection properties. 
Thus, a profile of finite range can be extracted which might be 
utilized by nanotechnology. Although we can not force the
profile to be limited to $x>0$, the procedure allows to create
non-vanishing profiles with a rather sharp edge at $x=0$. 

A severe problem for applications is the necessity to provide
scattering information in the energetically closed channels. In the
examples  presented here, we overcome this difficulty by using simulated 
data. However, for future applications a practically feasible and 
problem orientated method for analytical continuation has still to 
be developed. 

As already mentioned, the design of quantum devices via
nanotechnology seems to be a potential field of application of these
inverse scattering methods on the line (see also Ref. \cite{QD99}).
 In the presence of couplings between different electronic bands, 
the construction of profiles with 
specific reflection and/or transmission properties via fitting methods 
guided by intuition becomes rather difficult. Here, the developed
coupled-channel inverse scattering methods can give a reliable guide
line towards the required profile for the device. In this context the
inclusion of bound states in the continuum in the inversion 
scattering procedure would be of great interest because slight 
changes in the retrieved profiles may result in sharp resonances
which could be of interest for applications. Work is in progress
to handle this important question from the inverse scattering point
of view. 

The presented formalism can be applied to any technological problem
where one wants to retrieve or design the profile from the spectral
information. Nanostructure devices and waveguides are obvious 
examples where these procedures can be of great interest. 
\acknowledgments
The authors gratefully  acknowledge financial support from the
the University of South Africa and the Vienna University of 
Technology.

\appendix
\section{Derivation of The Marchenko Equation}
\label{app:A}
The formulation of the solution of the ISP by means of an integral 
equation, frequently referred to as the Marchenko equation, is most 
easily achieved via the Levin representation of the Jost solution 
{\bff}$_-(k,x)$ (cf. Eq. (\ref{Levin2})).
In a first step we multiply Eq. (\ref{physL}) from the right by
\begin{equation}
  \frac{1}{2\pi}\exp(-{\rm i}\mbox{\KK}y) 
  \frac{{\rm d}\mbox{\KK}}{{\rm d} k}
  \equiv \frac{1}{2\pi}\exp(-{\rm i}\mbox{\KK}y)\mbox{\KK}^{-1}k
\, ,
\label{derK}
\end{equation}
substitute the Levin representations for {\bff}$_-(k,x)$ and 
{\bbff}$_-(k,x)$ on the right hand side and integrate over $k$.
Writing {\TT}$_L=\um + ${\GG}$(k)$ and reordering the terms leads 
to the relationship
%
\begin{eqnarray}
       \mbox{\II}_1&+&\mbox{\II}_2 = \mbox{\bbfrho}_-(x,y)
\label{Mar01}
\\ &+&
       \int_{-\infty}^x {\rm d}z \mbox{\BB}_-(x,z) \mbox{\bbfrho}_-(z,y)
       \int_{-\infty}^x {\rm d}z \mbox{\BB}_-(x,z) \mbox{\HH}(z,y)
\nonumber
\end{eqnarray}
%
with
\begin{eqnarray*}
\mbox{\II}_1(x,y) & = & \frac{1}{2\pi} \int_{-\infty}^{+\infty} {\rm d}k k
       \mbox{\bff}_+(k,x)\mbox{\GG}(k){\rm e}^{-i\mbox{\KK}y}
       \mbox{\KK}^{-1}\\
\mbox{\II}_2(x,y) & = & \frac{1}{2\pi} \int_{-\infty}^{+\infty} {\rm d}k k
   [\mbox{\bff}_+(k,x)-{\rm e}^{-i\mbox{\KK}x}]{\rm e}^{-i\mbox{\KK}y}
   \mbox{\KK}^{-1}\\
   \mbox{\bbfrho}_-(x,y) & = & \frac{1}{2\pi} \int_{-\infty}^{+\infty} 
   {\rm d}k k {\rm e}^{-i\mbox{\KK}x} \mbox{\RR}_L(k)
   {\rm e}^{-i\mbox{\KK}y} \mbox{\KK}^{-1}\\
\mbox{\HH}(x,y) & = & \frac{1}{2\pi} \int_{-\infty}^{+\infty} {\rm d}k k
   {\rm e}^{-i\mbox{\KK}(x-y)} \mbox{\KK}^{-1} \, .
\end{eqnarray*}
In the presence of thresholds these integrals must be handled
consistently and with care because of the $(N-1)$-fold branch cut of 
the physical sheet of the Riemann surface for the momentum variable
$k$. Because of the identity (\ref{derK}) the j-th columns of the
matrices {\II}$_1$, {\II}$_2$, {\bbfrho}, {\HH} contain effectively 
integrals over the channel wave numbers $k_j$. It is therefore 
important to consider the consequences of the mapping of $k$ to $k_j$, 
as given in Eq. (\ref{kjj}). As displayed in Fig. 5
the integral 
over the real axis of $k$ is transformed into a contour extending
also to the positive imaginary axis of $k_j, j=2,\dots ,N$.
%
%
%

To show the effect of this deformation of the contour we consider
the matrix {\HH}$(x,y)$ which is diagonal with elements
\begin{equation}
       \mbox{\HH}_{jj}(x,y) = \frac{1}{2\pi}\int_{-\infty}^{+\infty} 
        {\rm d}k\frac{{\rm d}k_j}{{\rm d}k} {\rm e}^{ik_j(x-y)} 
\end{equation}
with $j=1,\cdots, N$, $k_{11}\equiv k$.
Performing a transformation from $k$ to $k_j$ in the $jj$-element
leads to a splitting of the integral
\begin{eqnarray}
     H_{jj}(x,y)&=&\frac{1}{2\pi } 
            \int_{-\infty}^{+\infty} {\rm d} k_j {\rm e}^{ik_j(x-y)}
\label{Hjj}
\\
   &-i& \frac{1}{2\pi}\int_{-\sqrt{\epsilon_j}}^{+\sqrt{\epsilon_j}}
          \dr k \frac{k}{\sqrt{\epsilon_j-k^2}} 
          {\rm e}^{-\sqrt{\epsilon_j-k^2}(x-y)} 
\nonumber
\end{eqnarray}
Because of the symmetry of the integrand the second term of Eq.
(\ref{Hjj}) vanishes, while the first term yields the $\delta$-function.
Hence we obtain {\HH}$(x,y)=\delta (x-y) \um $. 

Making use of the Levin representation of {\bff}$_-(x,y)$ we can 
apply the expression for {\HH}$(x,y)$ to evaluate {\II}$_2(x,y)$. 
Restricting ourselves to $x>y$, the matrix {\II}$_2(x,y)$ vanishes and
Eq. (\ref{Mar01}) takes the simple form
\begin{eqnarray}
\nonumber
        \mbox{\II}_1(x,y)&= &\mbox{\bbfrho}_-(x,y)
      + \mbox{\BB}_-(x,y) \Theta (x-y)
\\
       &+& \int_{-\infty}^x {\rm d}z \mbox{\BB}_-(x,z) \mbox{\bbfrho}_-(z,y)
\label{Mar02}
\end{eqnarray}
where $\Theta (x)$ is the Heavyside function.

In Eq. (\ref{Mar02}) we are only left with the evaluation of the 
integral {\II}$_1(x,y)$. To evaluate {\II}$_1(x,y)$ we close the 
integral in the upper-half plane of $k$ in order to apply Cauchy's 
theorem. To do so we have to consider the integrand with regard to its 
analyticity. It is well known from scattering theory \cite{Chadan89}
that the occurrence of bound states leads to a pole of {\TT}$(k)$
and {\RR}$(k)$ on the positive imaginary axis of $k$. This is best 
seen from Eq. (\ref{physL}) where the requirement of a normalizable 
solution at $k=q=i\kappa , \kappa >0$ requires the relationship
\begin{equation}  
\mbox{\bff}_+(i\kappa ,x) = \mbox{\bff}_-(i\kappa ,x)
\lim_{k\to i\kappa } \mbox{\RR}_L(k) \mbox{\TT}_L^{-1}(k)
\, .
\label{bound}
\end{equation}
Employing a method similar to the one used in \cite{Wadati74,Weidenm64},
it can be shown that the poles of {\TT}$_L(k)$ at 
$q_\alpha =i\kappa_\alpha $ are simple, i.e.
\begin{equation}
\mbox{\TT}_L(k)= \frac{1}{k-q_\alpha} \mbox{\NN}_\alpha
+ \cdots \, ,
\end{equation}
where {\NN}$_\alpha$ is the residue of {\TT}$_L$ at $k=q_\alpha $
and therefore also of {\GG}$_\alpha$.

In the following we restrict ourselves to systems which have only 
true bound states at $k^2=q_\alpha^2<0, \alpha =1, \dots ,N_b$.
Thus the Jost solution is well defined in the upper half-plane 
of each channel wave number $k_j$ and the integrand does not 
exhibit irregular features along the contour of integration. 
In order to evaluate the integral over the upper half circle
we must consider the behavior of the {\bff}$(k,x)$ for 
$|k|\to \infty$. From Eq. (\ref{kjj}) it immediately follows that
$$
\mbox{Im }k_j = \mbox{Im }k \left[1+\frac{1}{2}\frac{\epsilon_j}{|k|^2}+
o\left( \frac{\epsilon_j^2}{|k|^4} \right)\right]
\, .
$$
Hence, the dominant term for the vanishing of the Jost solution in the
upper half plane of $k$ for $|k|\to \infty$ is independent of the
channel and leads to a vanishing of the integral over the half circle
for $x>y$. Thus the integral {\II}$_1(x,y)$ is simply given by the sum
over all bound state poles
\begin{equation}
  \mbox{\II}_1=i \sum_{\alpha =1}^{N_b}\mbox {\bff}_+(q_\alpha ,x) 
  \mbox{\NN}_\alpha \exp(-i\mbox{\QQ}_\alpha y) \mbox{\QQ}(q_\alpha )
  ^{-1}q_\alpha \, ,
\end{equation}
where we assume that the potential {\VV}$(x)$ sustains $N_b$ 
bound states.

Using the specific relationship, Eq. (\ref{bound}), between the 
Jost functions {\bff}$_+$ and {\bff}$_-$ at the bound states, 
$k=q_\alpha $, and entering the Levin representation for
{\bff}$_-(k,x)$ leads finally to
\begin{equation}
  \mbox{\II}_1(x,y) = -\mbox{\bbfrho}^{(b)}_-(x,y) -
       \int_{-\infty}^{x}\, {\rm d}z \mbox{\BB}_-(x,z) 
       \mbox{\bbfrho}^{(b)}_-(z,y)
\end{equation}
with
\begin{equation}
  \mbox{\bbfrho}^{(b)}_-(x,y)=-\sum_{\alpha=1}^{N_b} 
         \exp(-i\mbox{\QQ}_\alpha x)
        i\mbox{\MM}_\alpha \exp(-i\mbox{\QQ}_\alpha y)
        \mbox{\QQ}_\alpha^{-1}q_\alpha
\end{equation}
where {\MM}$_\alpha $ are the residues of {\RR}$_L(k)$ at $q_\alpha $.
Introducing the total input kernel
\begin{equation}
\mbox{\bfrho}_-(x,y)=\mbox{\bbfrho}_-(x,y)+\mbox{\bbfrho}^{(b)}_-(x,y)
\end{equation}
leads to the Marchenko equation (\ref{Mar1m}).

\section{Definition of Potential Profiles}
\label{app:B}

In the examples we use potentials, whose elements
are given by an $x$-dependence of closed form. In the
following we summarize the definitions of the potential 
forms and their parameters.

\subsection{Gaussian Profile}
\begin{equation}
           V(x)=V_0\exp (-b(x-c)^2)
\label{eq:gauss}
\end{equation}

\bigskip

\subsection{Multilayer Profile}
\begin{eqnarray}
\nonumber
       V(x) &=& \sum_{i=1}^N V_i \left[ 
          \frac{1}{1+\exp ((x_{i-1} -x)/a)}\right. \\
& - &
        \left.   \frac{1}{1+ \exp ((x_i -x)/a)}
         \right] \, ,
\label{eq:multil}
\end{eqnarray}
where $N$ is the number of layers, $a$ is the diffuseness
parameter, and $V_i$ the strength of the layer with width
$x_i-x_{i-1}$ with $x_0$ being the left boundary of the 
multilayer profile.

\bigskip

\subsection{Multiple Sea-Saw Potential}
The sea-saw potential is defined by
\begin{equation}
        V(x)=\frac{2\ V_0}{x_\ell} 
\begin{cases}
          0 & x< x_0 \cr 
          s_m\ (x - x_{m-1}) & x_{m-1} \leq x \leq \bar x_m\cr
          s_m\ (x_m - x) & \bar x_m \leq x \leq x_m \cr
            & \phantom{abcdef} m=1,\dots ,n_s\cr
          0 & x_{n_s} < x \cr
\end{cases}
\, ,
\label{eq:seasaw}
\end{equation}
where $x_m=x_s+m x_\ell $ and $\bar x_m=x_s+(m-\frac{1}{2}) x_\ell$,
$m=1,\dots ,n_s$. Here, $n_s$ is the number of sea-saw peaks of length 
$x_\ell $, $V_0$ is the strength of the potential, and $s_m$ is the 
sign of the corresponding term. The nonvanishing part of the 
profile is shifted by an amount $x_s$ from the origin. Thus it extends 
between $x_0$ and $x_{n_s}$.

\vfill
\newpage

\section*{Figure Captions}

{\bf Figure 1}\\
Reconstructed potential (dashed line) of a 2-channel system
without thresholds or bound states. The details of the original 
potential (solid line) are given in the text.

\vspace{0.5cm}
{\bf Figure 2}\\
Reconstructed potential (solid line) of a coupled 2-channel
system with a threshold $\epsilon_2=0.025$ in the 22-channel. 
The details of the original potential (dashed line) are given in 
the text.

\vspace{0.5cm}
{\bf Figure 3}\\
Reconstructed potential (dashed line) of a two-channel 
system with a bound state but  without thresholds.
The original potential (solid line) is shown for comparison.

\vspace{0.5cm}
{\bf Figure 4}\\
Input potentials (solid line)  and the  
SUSY transformation potentials (dashed line).

\vspace{0.5cm}
{\bf Figure 5}\\
Mapping of $k$ to $k_j$ and the  corresponding deformation
of the contour for integration.



\newpage

\begin{center}

\begin{figure}[htb]
\centerline{\epsfig{file=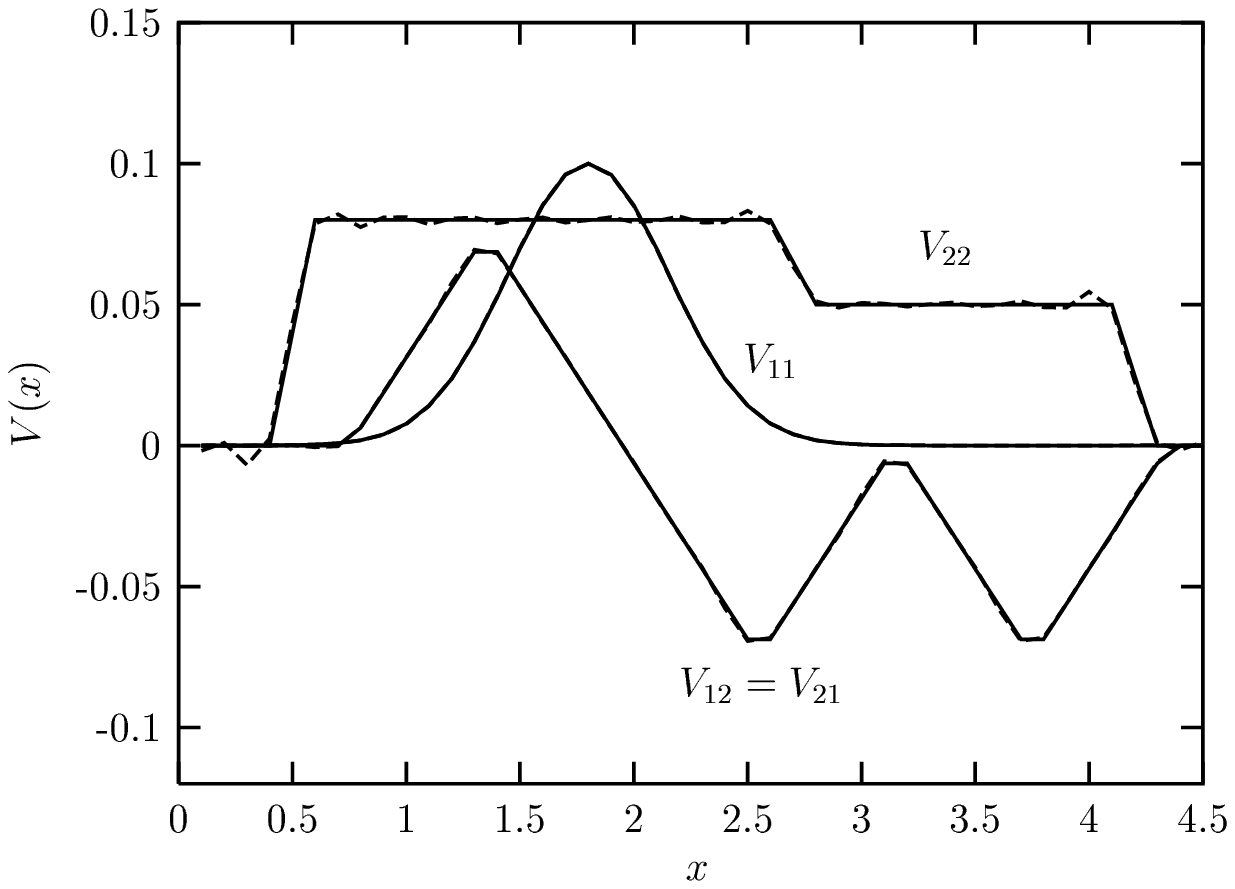,width=16.0cm}}
\label{fig1}
\end{figure}

\vspace{3cm}

{\large {\bf Figure 1}}

\newpage

\begin{figure}[htb]
\centerline{\epsfig{file=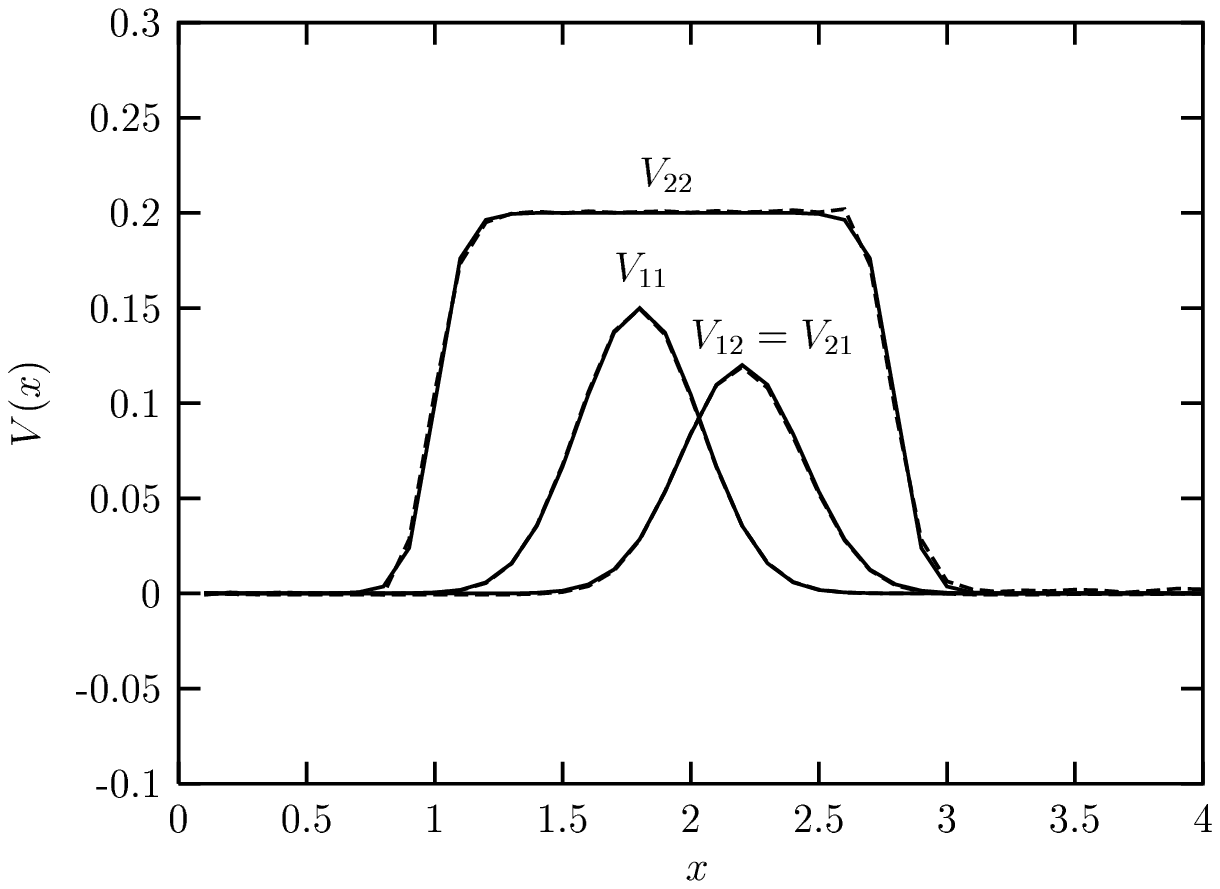,width=16.0cm}}
\label{fig2}
\end{figure}

\vspace{3cm}

{\large {\bf Figure 2}} 

\newpage

\begin{figure}[htb]
\centerline{\epsfig{file=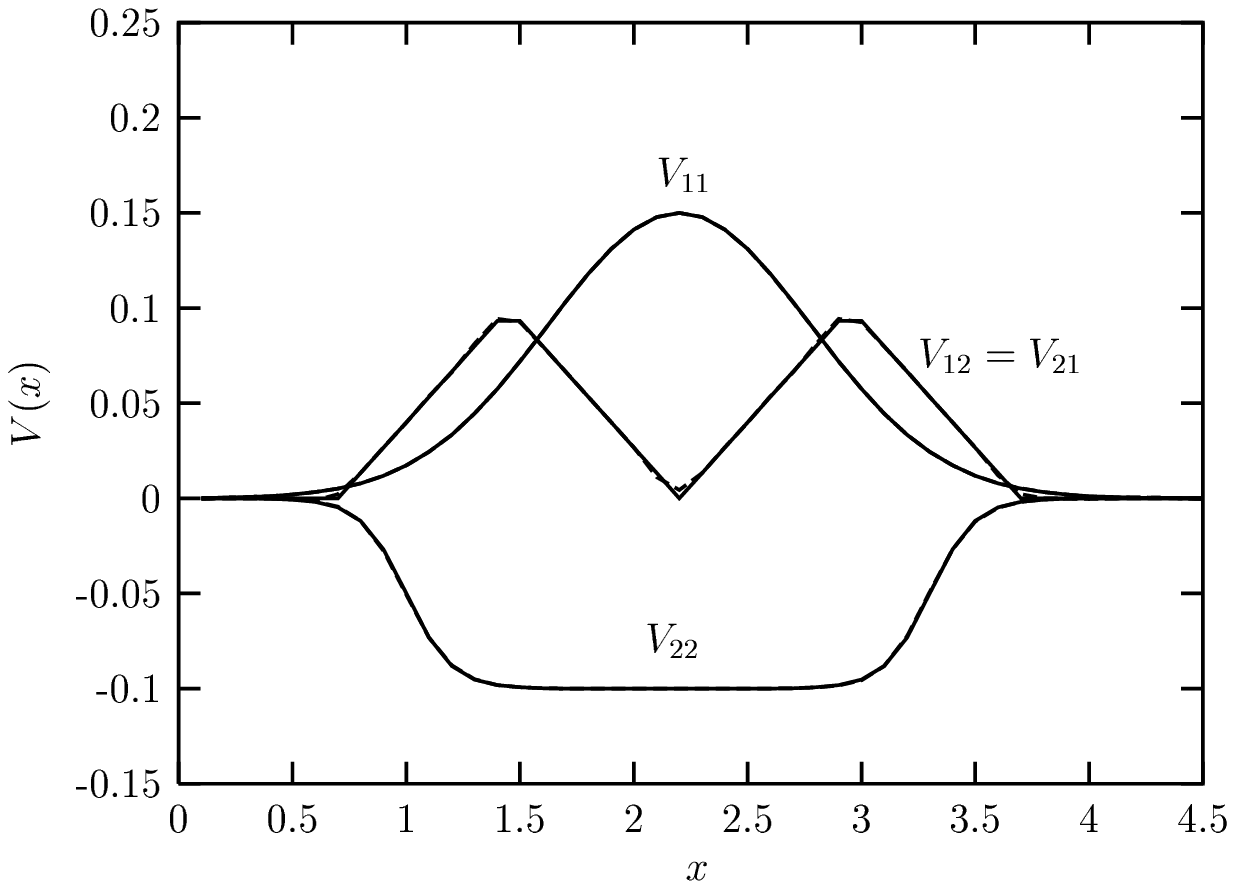,width=16.0cm}}
\label{fig3}
\end{figure}

\vspace{3cm}

{\large {\bf Figure 3}}

\newpage

\begin{figure}[htb]
\centerline{\epsfig{file=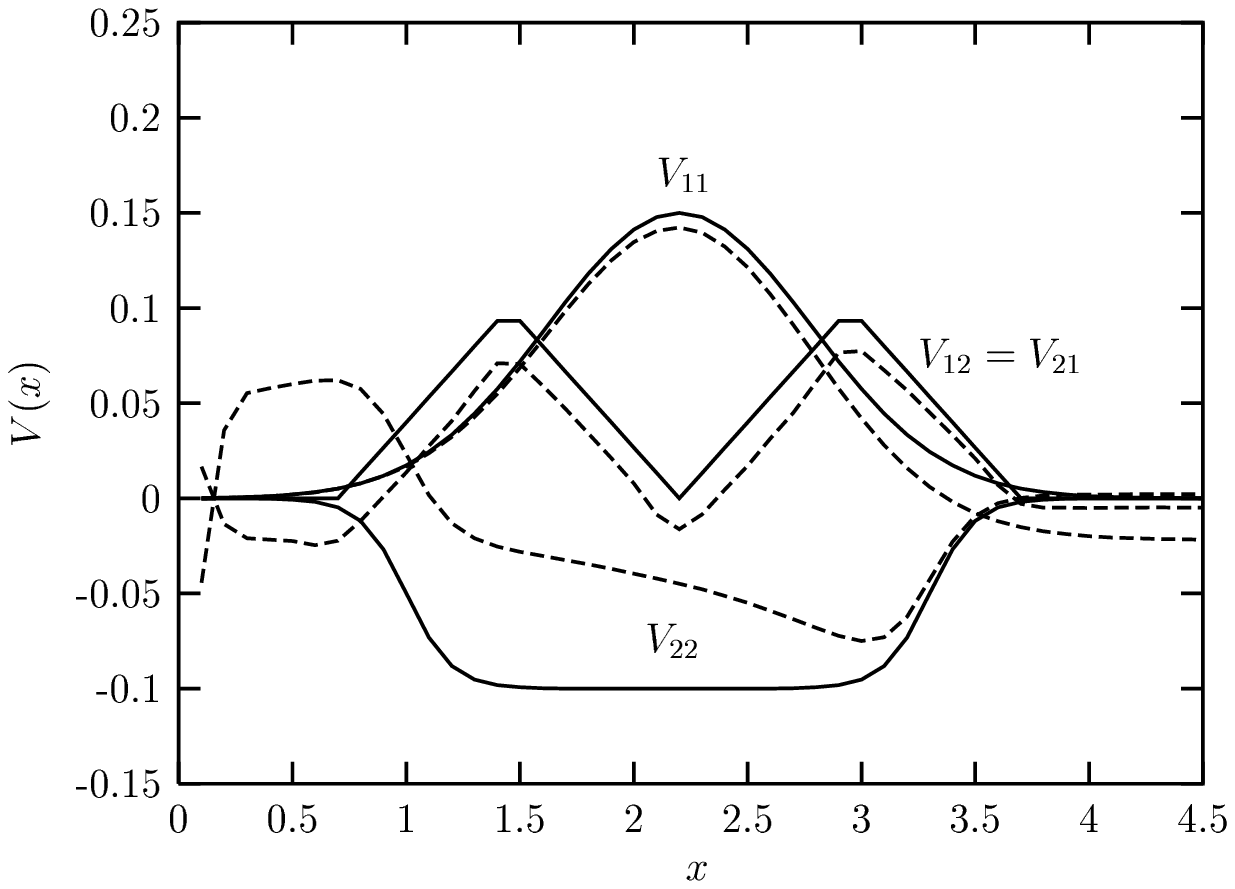,width=16.0cm}}
\label{fig4}
\end{figure}

\vspace{3cm}

{\large {\bf Figure 4}}

\newpage

\begin{figure}[htb]
\centerline{\epsfig{file=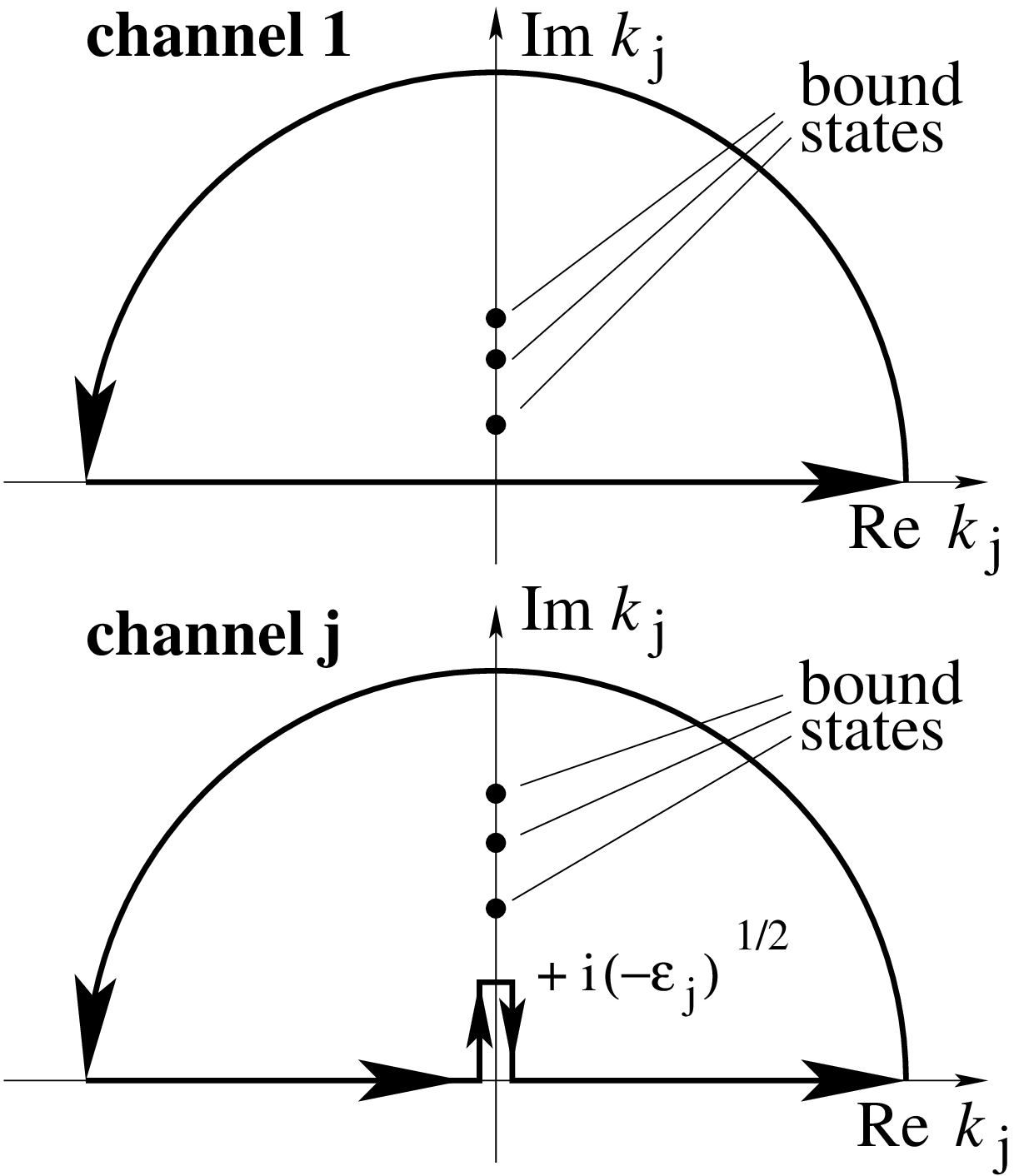,width=16.0cm}}
\label{Fig5}
\end{figure}

\vspace{3cm}

{\large {\bf Figure 5}}

\end{center}
%

\begin{thebibliography}{99}

\bibitem{Kay60}
        I. Kay,  Comm. Pure Appl. Math {\bf 13}, 371 (1960).
%
\bibitem{Marchenko63}
        Z. S.  Agranovich and V. A.  Marchenko,
        {\it The Inverse Problem of Scattering Theory},
        (Gordon \& Breach, New York, 1963).
%
\bibitem{Chadan89}
        K. Chadan and P. C. Sabatier,
        {\em Inverse Problems in Quantum Scattering Theory}, 2nd ed.
        (Springer, New York, 1989).
%
\bibitem{Lamb81}
        G. L. Lamb,
        {\em Elements of Soliton Theory} (Willey, New York), 1981.
%
\bibitem{Ghosh91}
        D.\,N. Ghosh Roy, {\it Methods of Inverse Problems
         in Physics} (CRC Press, Boston, 1991).
%
\bibitem{Wadati74}
     M. Wadati and  T. Kamijo T,  {\it Prog. Theoret. Phys.}
     {\bf 52},  397 (1974).
%
\bibitem {Calogero77}
     F. Calogero  and D.  Degasperis, Nuovo Cimento {\bf 32B}, 201
     (1976); {\bf 39B}, 1 (1977).
%
\bibitem{Sofianos97}
         S. A. Sofianos, M. Braun, R. Lipperheide, and H. Leeb,
         {\it Lecture Notes in Phys.} {\bf 488}, 54 (1997).
%
\bibitem{Cowley75}
        J. M. Cowley, 
        {\em Diffraction Physics} (North Holland, Amsterdam), 1975.
%
\bibitem{Burge76}
         R. E. Burge, M. A. Fiddy, A. H. Greenaway, and G. Ross,
         Proc. Roy. Soc. London {\bf A 350}, 191 (1976).
%
\bibitem{Klibanov92}
         M. V. Klibanov and P. E. Sacks, 
         J. Math.  Phys. {\bf 33}, 3813 (1992).
%
\bibitem{Kasper99}
         J. Kasper, H. Leeb and R. Lipperheide, J. Magn. Magn.
         Mat. {\bf 196}, 51 (1999).
%
\bibitem{QD99}
        Microelectronics Journal {\bf 30} (1999). The whole
        volume is devoted to various aspects of
        semiconductor quantum  devices.
%
\bibitem{Braun95}
        M. Braun, S. A. Sofianos, and R. Lipperheide, Inverse
        Problems {\bf 11}, L1 (1995).
%
\bibitem{Witten81}
        E. Witten, Nucl. Phys. B {\bf 188}, 51 (1981).
%
\bibitem{Baye87}
        D. Baye, Phys. Rev. Lett. {\bf 58}, 2738 (1987); 
                Phys. Rev. A {\bf 48}, 2040 (1993).
%
\bibitem{Sukumar85} 
        C. V. Sukumar, J. Phys. A: Maths. Gen. {\bf 18},
        L57 (1985); {\bf 18}, 2917 (1985); {\bf 18}, 2937 (1985).
%
\bibitem{Infeld51}
        L. Infeld and T. E. Hull, Rev. Mod. Phys. {\bf 23}, 21 (1951).
%
\bibitem{Amado87}
        R. D. Amado, F. Cannata, J.-P. Dedonder, Phys. Rev. Lett. 
        {\bf 61}, 2901 (1987); Int. J. Mod. Phys. A {\bf 5}, 
        3401 (1990).
%
\bibitem{Sparenberg97}
        J.-M. Sparenberg and D. Baye, Phys. Rev. Lett. {\bf 97},
        3802 (1997).
%
\bibitem{Newton}
        R. G. Newton,
        {\em Scattering Theory of Waves and Particles}, 2nd ed.
        (Springer, New York, 1982). 
\bibitem{Leeb00}
        H. Leeb, S. A. Sofianos, J.-M. Sparenberg, and D. Baye,
        Phys. Rev. {\bf C 62}, 064003 (2000).
%
\bibitem{Weidenm64}
        H. A. Weidenm\"uller, Ann. Phys. (N.Y.) {\bf 29}, 60 (1964).
\end{thebibliography}
\end{document}